\documentclass[twocolumn,english,aps,prl,superscriptaddress,amsmath,longbibliography]{revtex4-2}
\usepackage[T1]{fontenc}
\usepackage[latin9]{inputenc}
\usepackage{amsmath}
\usepackage{graphicx}
\usepackage{color}
\usepackage{float}
\usepackage{graphicx,color,hyperref,xcolor}
\makeatletter
\usepackage{babel}

\makeatother

\usepackage{babel}
\begin{document}
\title{Enhanced Cooper Pairing via Random Matrix Phonons in Superconducting
Grains}
\author{Andrey Grankin}
\affiliation{Joint Quantum Institute, Department of Physics, University of Maryland, College Park, MD 20742, USA}
\author{Mohammad Hafezi}
\affiliation{Joint Quantum Institute, Department of Physics, University of Maryland, College Park, MD 20742, USA}
\author{Victor Galitski}
\affiliation{Joint Quantum Institute, Department of Physics, University of Maryland, College Park, MD 20742, USA}
\begin{abstract}
There is rich experimental evidence that granular superconductors
and superconducting films often exhibit a higher transition temperature,
$T_{c}$, than that in bulk samples of the same material. This paper
suggests that this enhancement hinges on random matrix phonons mediating
Cooper pairing more efficiently than bulk phonons. We develop the
Eliashberg theory of superconductivity in chaotic grains, calculate
the random phonon spectrum and solve the Eliashberg equations numerically.
Self-averaging of the effective electron-phonon coupling constant
is noted, which allows us to fit the numerical data with analytical
results based on a generalization of the Berry conjecture. The key
insight is that the phonon density of states, and hence $T_{c}$,
shows an enhancement proportional to the ratio of the perimeter and
area of the grain - the Weyl law. We benchmark our results for aluminum
films, and find an enhancement of $T_{c}$ of about $10\%$ for a
randomly-generated shape. A larger enhancement of $T_{c}$ is readily
possible by optimizing grain geometries. We conclude by noticing that
mesoscopic shape fluctuations in realistic granular structures should
give rise to a further enhancement of global $T_{c}$ due to the formation
of a percolating Josephson network. 
\end{abstract}
\maketitle
The Bardeen-Cooper-Schrieffer (BCS) theory of superconductivity is
a rare example of a controlled theory with a quantitative relevance
to experiment. It has been tremendously successful not only in explaining the origin of superconductivity, but also in accurately estimating the transition temperature in a variety of conventional phonon-driven superconductors. However, despite this success, there exists an extensive range of experimental phenomenology on granular superconductors, disordered films, and layered structures dating from the 1940s up to these days that remains largely unexplained \cite{abeles1966enhancement,thomas2019exploring,smolyaninova2016enhanced,cohen1968superconductivity,strongin1968enhanced,prischepa2023phonon}. Paradoxically, it has been observed that making superconducting structures more random and granular often leads to an increase of the
superconducting transition temperature, $T_{c}$, sometimes exhibiting
many-fold increase \cite{abeles1966enhancement} of $T_{c}$ compared to bulk three-dimensional
samples. Unfortunately, the standard computational material science
techniques are not informative in this context, because the underlying 
band theory breaks down.

This work  develops the theory of superconducting pairing in mesoscopic grains. 
A generic grain boundary defines a chaotic billiard, and therefore both the electron \cite{garcia2011bcs,garcia2008bardeen}
and phonon spectra follow random matrix theory \cite{tanner2007wave}. These spectra are eigenvalues of the 
 elliptic differential operators originating -- the single-particle Schr{\"o}dinger 
equation for electrons and the wave equation for phonons. In the simplest case of an elementary
metal (e.g., aluminum) only acoustic modes are relevant \cite{animalu1966phonon}. The Debye model further reduces the problem to solving the Laplace equation for transverse and longitudinal phonons, which are coupled through non-trivial boundary conditions, $n_i \sigma_{ij}\Bigr|_{\rm boundary} = 0$ \cite{tanner2007wave,achenbach1982ray,landau2012theory},
where ${\bf n}$ is normal to the boundary and $\hat{\sigma}$ is the stress tensor defined below.
The properties of the spectra of Laplace operators as a function of geometry and type of the 
boundary is an old question going back to the 1911 work by Weyl \cite{weyl1911asymptotische}. For Neumann-type boundary conditions, 
which are the case for a phonon billiard, the Weyl law establishes a {\em positive} mesoscopic correction to the density of states (DoS) proportional to the ratio of the perimeter, $P$, and area, $A$ of the billiard \cite{bertelsen2000distribution,tanner2007wave}. Specifically, for acoustic phonons in two dimensions, the correction to the total DoS is
\begin{equation}
\frac{1}{A} \left[ N(\omega)-N_{\text{Bulk}}(\omega) \right] = \frac{\eta\omega}{4\pi c_{\perp}}\frac{P}{A}  +o (\omega),\label{eq:Nw}
\end{equation}
where $N\left(\omega\right)=   \sum_{\omega_{l}\leq\omega}1$ is the number of eigenvalues $\omega_l < \omega$, $c_{\perp}$ ($c_{||}$) is the velocity of transverse (longitundinal) phonons,  and $\eta$ is a positive dimensionless constant  of order one,  which for free-surface boundary conditions depends on the ratio $c_{\perp}/c_{||}$ only \cite{bertelsen2000distribution} (see appendix and Fig.~\ref{Fig1}). The longitudinal phonon DoS also acquires
a positive correction (see, Fig.~\ref{Fig1}b), which eventually
translates into an enhancement of the superconducting transition temperature. 
\begin{equation}
\frac{T_{c}-T_{c0}}{T_{c0}}\propto \frac{\delta\lambda}{\lambda_{\rm{bulk}}} \approx\frac{\eta_{\parallel}P}{2Ak_{D}} \log\left(\frac{\omega_{\text{D}}}{\omega_{0}e^{\chi}}\right) -\frac{\omega_{0}}{\omega_{\text{D}}},\label{Tc}
\end{equation}
where $k_{D}=\omega_{D}/c_{||}\sim a^{-1}$ is the Debye cutoff of
order inverse lattice constant, whose exact value is determined by
the total number of available phonon modes held constant for a given
area, $A$.  In Eq.~(\ref{Tc}), we assumed that the Fermi wavelength is the smallest lengthscale. {$\lambda_{\rm{bulk}}$ and $\delta\lambda$ respectively denote the bulk BCS coupling strength and its modification in our geometry. $\eta_{\parallel}\sim 1$ is a dimensionless constant plotted in the inset of Fig.~\ref{Fig1}b and $\omega_{0}~\sim c_{\parallel}\pi/\sqrt{A}$ is a non-universal low-energy cut-off of order finite-size quantization energy and $\chi=\log\frac{c_{\parallel}c_{\perp}}{\left(c_{\parallel}^{2}+c_{\perp}^{2}\right)}\frac{\eta}{\eta_{\parallel}}$. 

\begin{figure*}[hbt!]
\centering \includegraphics[scale=0.7]{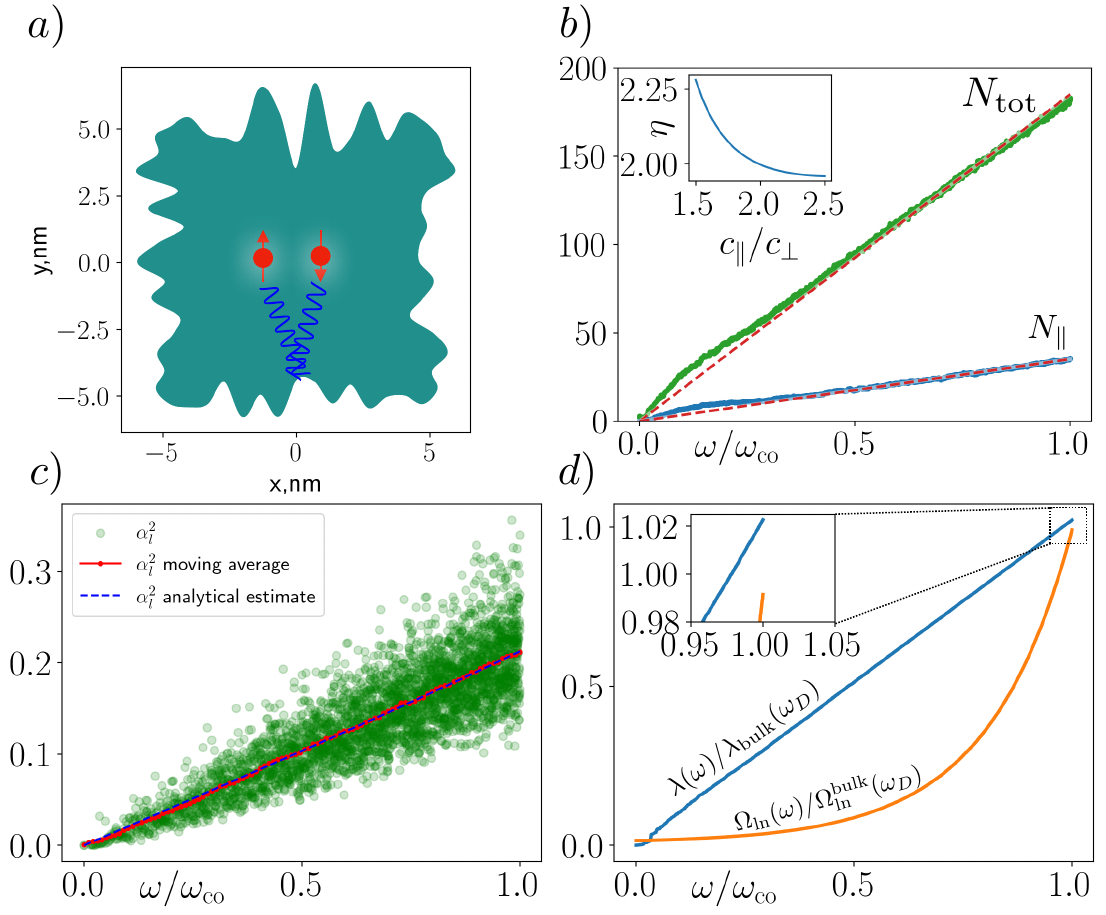} \caption{Electron pairing enhancement in chaotic grains. a)~Schematic illustration
of electron pairing in an irregular-shape metallic grain. b)~The total and the longitudinal contributions to the phonon density of states, $\delta N\left(\omega\right)=N(\omega)-N_{{\rm {bulk}}}(\omega)$, are
shown in green and blue respectively. Red dashed lines are linear fits to the data including
Eq.~\eqref{eq:Nw} for the total DoS. c)~The Fermi surface averaging of the overlap of the phonon eigenfuctions, $\alpha_{l}$, see Eq.~(\eqref{eq:Alpha}), as a function of the eigenstate energy. d)~Numerical results for the frequency-dependent BCS parameter $\lambda\left(\omega\right)\equiv2\int_{0}^{\omega}d\omega'\omega'^{-1}\alpha^{2}F\left(\omega'\right)$
and the logarithmic frequency cutoff $\Omega_{\text{ln}}\left(\omega\right)\equiv\omega_{\text{D}}\exp\left\{ \frac{2}{\lambda}\int_{0}^{\omega}d\omega'\omega'^{-1}\ln\left(\omega'/\omega_{\text{D}}\right)\alpha^{2}F\left(\omega'\right)\right\} $.
The superconducting transition temperature is determined by $\lambda=\lambda\left(\omega_{\text{co}}\right)$
and $\Omega_{\text{ln}}=\Omega_{\text{ln}}\left(\omega_{\text{co}}\right)$.}
\label{Fig1} 
\end{figure*}
 
%   We note that Eq.~(\ref{eq:Nw}) encodes the full
% DOS including transverse and longitudinal movement of lattice displacements
% which mix due to the boundary scattering {[}{]}. However, we can numerically
% separate the two contributions by projecting the eigenfunctions of
% the elastic equations onto the momentum direction. To this end, we
% solve the Navier-Cauchy equations in a grain geometry shown in Fig.~\ref{Fig1}~(a)
% with the free-surface boundary conditions. The resulting total and
% projected densities of states are also enhanced as shown in Fig.~\ref{Fig1}~(b).
% In this work we investigate how the modification of the density of
% states affects the electron pairing in chaotic grains. We want to
% emphasize that although the global density of states is enhanced,
% the total number of states has to be conserved. In this work we enforce
% this condition by using a cut-off energy $\omega_{\text{co}}$ that
% satisfies $N\left(\omega_{\text{co}}\right)=N_{\text{Bulk}}\left(\omega_{\text{D}}\right)$,
% where $\omega_{\text{D}}$ is Debye energy in bulk.

We consider a specific randomly generated shape shown in Fig.~\ref{Fig1}a, which is clearly chaotic. 
All numerical results are derived for this particular realization of a 2D grain, but due to self-averaging of relevant Eliashberg parameters, we are able to validate our specific computational data with generic analytical results rooted in random matrix theory.  The starting point is the following electron-phonon Hamiltonian

% \begin{figure}
% \includegraphics[scale=0.45]{Fig1}
% \caption{Electron pairing enhancement in chaotic grains. a) Schematic illustration
% of electron pairing in an irregular-shape metallic grain. b) Phononic
% cumulative state numbers $\delta N\left(\omega\right)=N(\omega)-N_{\rm{bulk}}(\omega)$ in
% chaotic grain. The total and the longitudinal contributions are shown in green and blue respectively. Red dashed lines represent an analytical expression (Eq.~) for the total and a fit for the longitudinal. c) Scaling of matrix element overlaps as a function of energy. d) Enhancement
% of BCS pairing strength $\lambda$. }
% \label{Fig1} 
% \end{figure}

\begin{align}
H & =\sum_{\sigma,m}\xi_{m}c_{m,\sigma}^{\dagger}c_{m,\sigma}+\sum_{\ell}\omega_{\ell}a_{\ell}^{\dagger}a_{\ell}\nonumber \\
 & +g\sum_{\sigma}\int_{A}d^{2}{\bf r}\nabla\cdot\vec{\Phi}\left({\bf r}\right)\psi_{\sigma}^{\dagger}\left({\bf r}\right)\psi_{\sigma}\left({\bf r}\right),\label{eq:H}
\end{align}
where  $\vec{\Phi}\left({\bf r}\right)=\sum_{\ell}\frac{\vec{\phi}_{\ell}\left({\bf r}\right)}{\sqrt{2\omega_{l}}}\left(a_{\ell}+a_{\ell}^{\dagger}\right)$
and $\psi_{\sigma=\uparrow,\downarrow}\left({\bf r}\right)=\sum_{m}\zeta_{m}\left({\bf r}\right)c_{m,\sigma}$ are the phonon and electron operators respectively and $g$ is the electron-phonon coupling. $\xi_m$ and $\zeta_{m}({\bf r})$ are the electron energies and wave-functions -- the spectrum of the Schr{\"o}dinger operator. Note  that in real materials, the mean free path for electrons is often much smaller  than the system size, $l \ll \sqrt{A}$ and hence random matrix theory description for electrons arises irrespective of boundary conditions.  In contrast, $\omega_{\ell}$ and $\vec{\phi}_{l}\left({\bf r}\right)$ are the phonon eigenfrequencies and eigenfunctions, which are sensitive to the boundary and follow from the Navier-Cauchy equation \cite{tanner2007wave} below 
\begin{equation}
\rho \, \ddot{\vec{\phi_{l}}} \equiv - \rho \omega_{\ell}^2 \vec{\phi_{l}} =\left(\xi+\nu\right) \nabla\left(\nabla\cdot\vec{\phi_{l}}\right)+\nu \Delta\vec{\phi_{l}},\label{eq:NC}
\end{equation}
where $\xi$ and $\nu$ are Lam\'e parameters, $\rho$ is the
material density, and the two sound velocities are $c_{\parallel}=\sqrt{\left(\xi+2\nu\right)/\rho}$, $c_{\perp}=\sqrt{\nu/\rho}$. We assume  $c_{\parallel}/c_{\perp}=2$ \cite{fassbender1989efficient,david1963attenuation,lide2004crc} and free-surface boundary condition.  
For a given grain geometry, Eq.~(\ref{eq:NC}) is solved using the finite-element methods available
in open-source software \cite{barrata2023dolfinx}. The total bulk number of states
below a certain energy $\omega$  is $N_{\text{Bulk}}\left(\omega\right)=\omega^{2}(c_{\parallel}^{-2}+c_{\perp}^{-2})/4\pi$.

We now generalize the Eliashberg theory  of phonon-mediated Cooper pairing to chaotic grains. Define electronic Nambu spinor fields $\Psi\left({\bf r}\right)=\{ \psi_{\uparrow}\left({\bf r}\right),\psi_{\downarrow}^{\dagger}\left({\bf r}\right)\} $
and the corresponding imaginary-time Green's function $\hat{{\cal G}}_{{\bf r},{\bf r}'}\left(\tau\right)=-\left\langle T\Psi\left({\bf r},\tau\right)\otimes\Psi^{\dagger}\left({\bf r}',0\right)\right\rangle $,
where $T$ is the time-ordering operator. The Nambu matrix-valued
self-energy is given by: $\hat{\Sigma}_{{\bf r},{\bf r}'}\left(\tau\right)=-g^{2}{\cal D}_{{\bf r},{\bf r}'}\left(\tau\right)\hat{\tau}_{3}\hat{{\cal G}}_{{\bf r},{\bf r}'}\left(\tau\right)\hat{\tau}_{3}$
where $\hat{\tau}_{i}$ are Pauli matrices in Nambu space and ${\cal D}_{{\bf r},{\bf r}'}\left(\tau\right)=-\left\langle T\left\{ \nabla\cdot\vec{\Phi}\left({\bf r},\tau\right)\nabla\cdot\vec{\Phi}\left({\bf r}',0\right)\right\} \right\rangle $. {Furthermore, we include electronic disorder by means of
an additional self-energy term $\hat{\Sigma}_{{\bf r}}=\left(4\pi\nu_{0}^{F}\tau_{\text{el}}\right)^{-1}\hat{\tau}_{3}\hat{{\cal G}}_{{\bf r},{\bf r}}\left(\tau\right)\hat{\tau}_{3}$,
where $\nu_{0}^{F}$ is the electronic DoS at the Fermi energy and
$\tau_{\text{el}}$ is the elastic scattering time. As we show in
the SM in diffusive limit, the superconducting gap obeys the following
local self-consistency equation:}
\begin{equation}
\Delta\left(i\epsilon_{n}\right)=-g^{2}\nu_{0}^{F}T\pi\sum_{n'}{\cal D}_{\text{eff}}\left(i\epsilon_{n}-i\epsilon_{n'}\right)\frac{\Delta\left(i\epsilon_{n'}\right)}{\left|\epsilon_{n}\right|Z\left(i\epsilon_{n'}\right)},\label{eq:Delta}
\end{equation}
{where $T$ is the temperature, $\epsilon_n=(2n+1)\pi T$ and $Z$ is the quasiparticle renormalization factor. The effective phonon propagator is defined by:}

\begin{equation}
{\cal D}_{\text{eff}}\left(i\Omega_{n}\right)=A^{-1}\int d^{2}{\bf r}d^{2}{\bf r}'J_{0}^{2}\left(k_{F}\left|{\bf r}-{\bf r}'\right|\right){\cal D}_{{\bf r},{\bf r}'}\left(i\Omega_{n}\right),\label{eq:Average propagator}
\end{equation}
where $J_{0}$ is the Bessel's function of the first kind, $k_{F}$
is Fermi momentum and $\Omega_n=2\pi nT$. Eqs.~(\ref{eq:Delta},~\ref{eq:Average propagator})
constitute the standard frequency-space Eliashberg equations \cite{marsiglio2020eliashberg}
and we can thus estimate the critical temperature using McMillan-Allen-Dynes
formula \cite{mcmillan1968transition,allen1975transition}: 
\begin{align}
T_{c}=\frac{\Omega_{\text{ln}}}{1.2}\exp\left\{ -\frac{1.04\left(\lambda+1\right)}{\lambda-\mu^{*}\left(1+0.62\lambda\right)}\right\}.\label{eq:MAD}
\end{align}
Here, the effective BCS coupling strength and the logarithmic cut-off frequency are defined as $\lambda=2\int d\omega\omega^{-1}\alpha^{2}F\left(\omega\right)$, $\Omega_{\text{ln}}=\omega_{\text{D}}\exp\left\{ \frac{2}{\lambda}\int d\omega\omega^{-1}\ln\left(\omega/\omega_{\text{D}}\right)\alpha^{2}F\left(\omega\right)\right\} $,
where $\omega_{\text{D}}$ is Debye energy, $\mu^{*}$ denotes
Coulomb pseudo-potential which we set to $0.1$ throughout and $\alpha^2 F(\omega)$ is the Eliashberg function~\cite{ummarino2013eliashberg,marsiglio2020eliashberg}:
\begin{align}
\alpha^{2}F\left(\omega\right) & =\frac{g^{2}\nu_{0}^{F}}{2A\omega}\sum_{l}{\delta\left(\omega-\omega_{l}\right)\alpha_{l}^{2}}\label{eq:Eliashberg}
\end{align}
where $\nu_{\text{tot}}^{\text{F}}$ is the electron density
of states and the dimensionless matrix element of phonon eigenstates averaged of Fermi surface  is  
\begin{align}
 \alpha_{l}^{2}=& \int d^{2}{\bf r}d^{2}{\bf r}'C_{l}\left({\bf r}\right)C_{l}\left({\bf r}'\right)J_{0}^{2}\left(k_{F}\left|{\bf r}-{\bf r}'\right|\right) ,\label{eq:Alpha}
\end{align}
where $\nu^{F}_0$ is the electronic DoS at the Fermi energy and we defined the divergence as $C_{l}\left({\bf r}\right)\equiv\nabla\cdot\vec{\phi}_{l}\left({\bf r}\right)$. 
%\textcolor{red}{We note that in clean limit the the electronic DoS also acquires corrections according to Weyl's law. }

We now provide a numerical estimation of the critical temperature
in the chaotic grain shown in Fig.~\ref{Fig1}~(a). Throughout this work we consider the limit $\xi\gg L\sim\sqrt{A}$, $k_F L\gg1$, $L\gg l$, where $l$ is the mean-free path and $\xi$ is the superconducting coherence length in diffusive limit (see SM). Assuming the following parameters for Aluminum films
$T_{c}^{\text{bulk}}=1.2K$ \cite{abeles1966enhancement} and $\omega_{\text{D}}/k_{\text{B}}\approx428K$
we get the bulk parameters $\Omega_{\text{ln}}^{\text{bulk}}/\omega_{\text{D}}\approx0.37$
and $\lambda_{\text{bulk}}\approx0.44$. We note that within our model,
we do not need the explicit knowledge of microscopic parameters such
as $g$.  For the chaotic grain in Fig.~\ref{Fig1}~(a),  the proper cutoff frequency  $\omega_{\text{co}}/\omega_{\text{D}}\approx0.97$ for
$c_{\perp}=3000m/s$. We now evaluate the matrix elements in Eq.~\eqref{eq:Alpha}
numerically (see Fig.~\eqref{Fig1}~(c)) and find the modified BCS
parameters $\lambda/\lambda_{\text{bulk}}\approx1.023$ and $\Omega_{\text{ln}}/\Omega_{\text{ln}}^{\text{bulk}}\approx0.99$.
Combining these factors, we get the critical temperature enhancement
$T_{c}/T_{c}^{\text{bulk}}\approx1.1$. 
\begin{figure}
\includegraphics[scale=0.31]{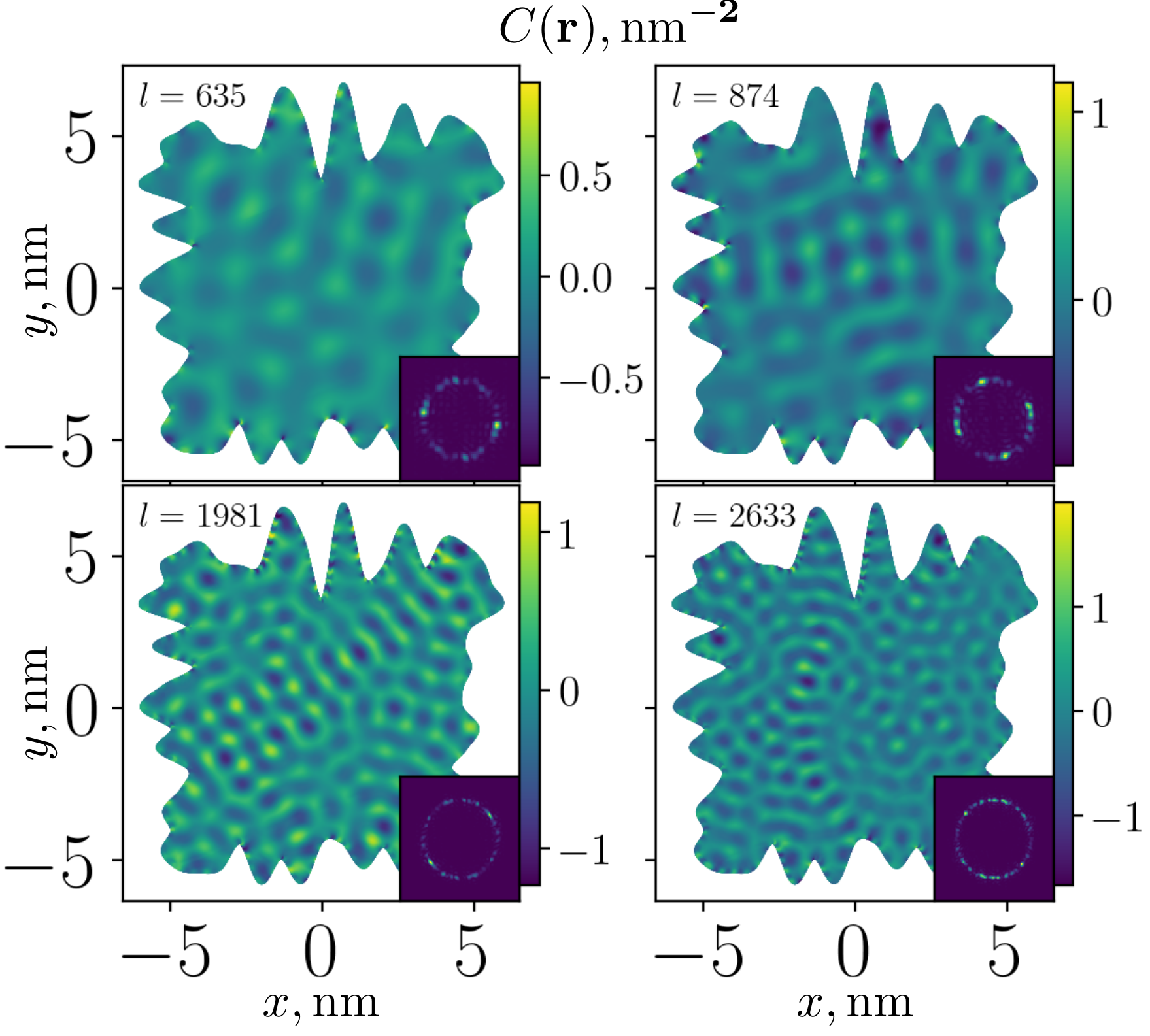}

\caption{Divergencies of eigenvectors of Navier-Cauchy equations in chaotic grain. Insets show their Fourier transforms $|C(\bf{k})|^2$ for $k_x,k_y\in[-1.6 k_l,1.6 k_l]$, where $k_{l}=\omega_{l}/c_{\parallel}$. Fourier transform is defined as $C\left({\bf k}\right)=\int d^{2}{\bf r}C\left({\bf r}\right)e^{-i{\bf kr}}$. }

\label{Fig2} 
\end{figure}

To get further insight, we apply random matrix theory to phonon eigenvectors. Typical divergences of eigenvectors $C_{l}\left({\bf r}\right)$ and their Fourier transforms at high
energies are shown in Fig.~(\ref{Fig2}). In momentum space, we observe a random-speckle structure at momenta corresponding to the energy of the state. Relying on the arguments pioneered by Berry \cite{berry1988semiclassical,tanner2007wave,akolzin2004generalized}, we conjecture that at
sufficiently short distances and sufficiently far away from the boundary,
the  correlation function of the phonon modes at high energies takes the following form (see also SM):

\begin{equation}
\overline{C_{l}\left({\bf r}\right)C_{l}\left({\bf r}'\right)}\approx A^{-1}\frac{\nu_{\parallel}^{\text{ph}}\left(\omega_{l}\right)}{\nu_{\text{tot}}^{\text{ph}}\left(\omega_{l}\right)}\frac{\omega_{l}^{2}}{c_{\parallel}^{2}}J_{0}\left(\frac{\omega_{l}}{c_{\parallel}}\left|{\bf r}-{\bf r}'\right|\right),\label{eq:phonon_correlation}
\end{equation}
where $\nu_{\parallel}^{\text{ph}}(\omega) = N'_{\parallel}(\omega)$ corresponds to the DoS of longitudinal phonons, which is $\nu_{\parallel}^{\text{ph}}\left(\omega\right)\equiv A^{-1}\left(\omega/c_{\parallel}\right)^{-2}\int d^{2}{\bf r}\sum_{l}C_{l}\left({\bf r}\right)C_{l}\left({\bf r}\right)\delta\left(\omega-\omega_{l}\right)$.
The resulting longitudinal DoS is shown in Fig.~(\ref{Fig2})~(c),
where we subtract the bulk contribution. We find that at high energies the DoS
$\nu_{\parallel}$  follows Weyl's law $\delta \nu_\parallel\approx \eta_\parallel P/(4\pi A c_\parallel)$ with  $\eta_\parallel\approx 0.79$. With this scaling we can also benchmark our assumption in Eq.~(\ref{eq:phonon_correlation}).
In Fig.~(\ref{Fig1})~(c) we plot the matrix elements $\alpha^{2}$
for different eigenstates and compare with the analytical formulas~(\ref{eq:phonon_correlation}, \ref{eq:Alpha}) $\alpha^{2}\left(\omega\right)\approx\{\nu_{\parallel}^{\text{ph}}\left(\omega\right)/\nu_{\text{tot}}^{\text{ph}}\left(\omega\right)\}4\omega/\sqrt{4c_{\parallel}^{2}k_{F}^{2}-\omega^{2}}$,
where  the total DoS $\nu_{\text{tot}}^{\text{ph}}(\omega) = N'(\omega)$. Together with Eqs.~(\ref{eq:Eliashberg}, \ref{eq:Alpha})
this yields the expression for the Eliashberg function at high energies:

\begin{align}
\frac{\alpha^{2}F\left(\omega\right)}{2 g^{2}\nu_{0}^{F}}\approx\frac{\nu_{\parallel}^{\text{ph}}\left(\omega\right)}{\sqrt{4c_{\parallel}^{2}k_{F}^{2}-\omega^{2}}},\label{eq:analytics}
\end{align}

Eq.~(\ref{eq:analytics}) implies log-singular corrections to the electron-phonon interaction parameter
$\lambda$, since the density of states is non-vanishing at low energies
according to Weyl's law Eq.~(\ref{eq:Nw}). However, our treatment
is valid only at sufficiently large energies and therefore we will
have to impose a  low-energy finite-size  cut-off $\omega_0$ which is sensitive to grain geometry and can be found by fitting to  numerical data. By doing so, we find  $\omega_0\sim\pi c_\parallel/\sqrt{A}$ for longitudinal phonons. 

We note that the enhancement of $T_c$ is a direct consequence of the Weyl's law which implies softening of phonons associated with the slower scaling of the density of states. Combining Eq.~(\ref{eq:analytics}), and taking into account the high- and low-energy cut-offs we get the analytical estimate for the $T_c$ enhancement provided in Eq.~\eqref{Tc} which, in our parameter regime, comes predominantly from the BCS coupling strength renormalization. We note that the low-energy cut-off is non-universal and can potentially be different for other grain geometries. 
% \textcolor{red}{Scaling argument? Debye model has no intrinsic scale }
%Finally, we note that  Eq.~\eqref{Tc} is expected to be correct for a wide range of grain sizes. This is guaranteed by the fact that Debye model has no intrinsic lengthscale except for the high-energy cut-off momentum. As a result, the and the whole grain spectrum can be scaled while preserving the shape.

In conclusion we note the existence of an optimal size of a grain for a given shape. Fixing the dimensionless parameter, $\zeta=P/\sqrt{A}$, and varying $T_c$~\eqref{Tc} over the characteristic size, $\sqrt{A}$,  we find the optimal condition as follows
$$
1=\frac{\omega_{0}}{\omega_{D}}\exp\left ({\frac{2A}{P\eta_{\parallel}}\frac{\omega_{0}}{c_{\parallel}}+\frac{c_{\perp}c_{\parallel}}{c_{\perp}^{2}+c_{\parallel}^{2}}\frac{\eta}{\eta_{\parallel}}+\eta_{\parallel}}\right )
$$
The corresponding change in the BCS strength is given by:
$\delta\lambda/\lambda_{\rm{bulk}}\approx\frac{\eta_{\parallel}\zeta}{2\pi}\exp\left\{ -\frac{2\pi}{\eta_\parallel\zeta}-\frac{c_{\perp}c_{\parallel}}{c_{\perp}^{2}+c_{\parallel}^{2}}\frac{\eta}{\eta_{\parallel}}-1\right\}$. We thus find that the critical temperature have a strong dependence on the grain geometry. For example for a circular grain $\zeta_\circ = 2/\sqrt{\pi}$, while for our grain $\zeta \approx 6.4$, which is significantly larger. Clearly, the geometry can be further optimized to create grains with a higher $T_c$ for a given material. 

Real granular superconductors are composed of a variety of grains of different shapes and sizes. Each grain has its own $T_c$ and there is a probability distribution of transition temperatures in the material. The overall positive shift -- Eq.~\eqref{Tc} -- corresponds to the average transition temperature $\langle T_c \rangle$. However, even for $T> \langle T_c \rangle$, there will exist a subset of grains with higher individual transition temperature. The global superconducting critical point in such a system is determined by a percolation transition where a Josephson network  of coupled superconducting grains spanning the entire sample first appears. This temperature can be considerably higher than $\langle T_c \rangle$, and such mesoscopic grain fluctuations provide another mechanism for enhancing superconductivity in granular materials similar to \cite{galitski2001disorder,galitski2008mesoscopic}.

\begin{figure}
\includegraphics[scale=0.3]{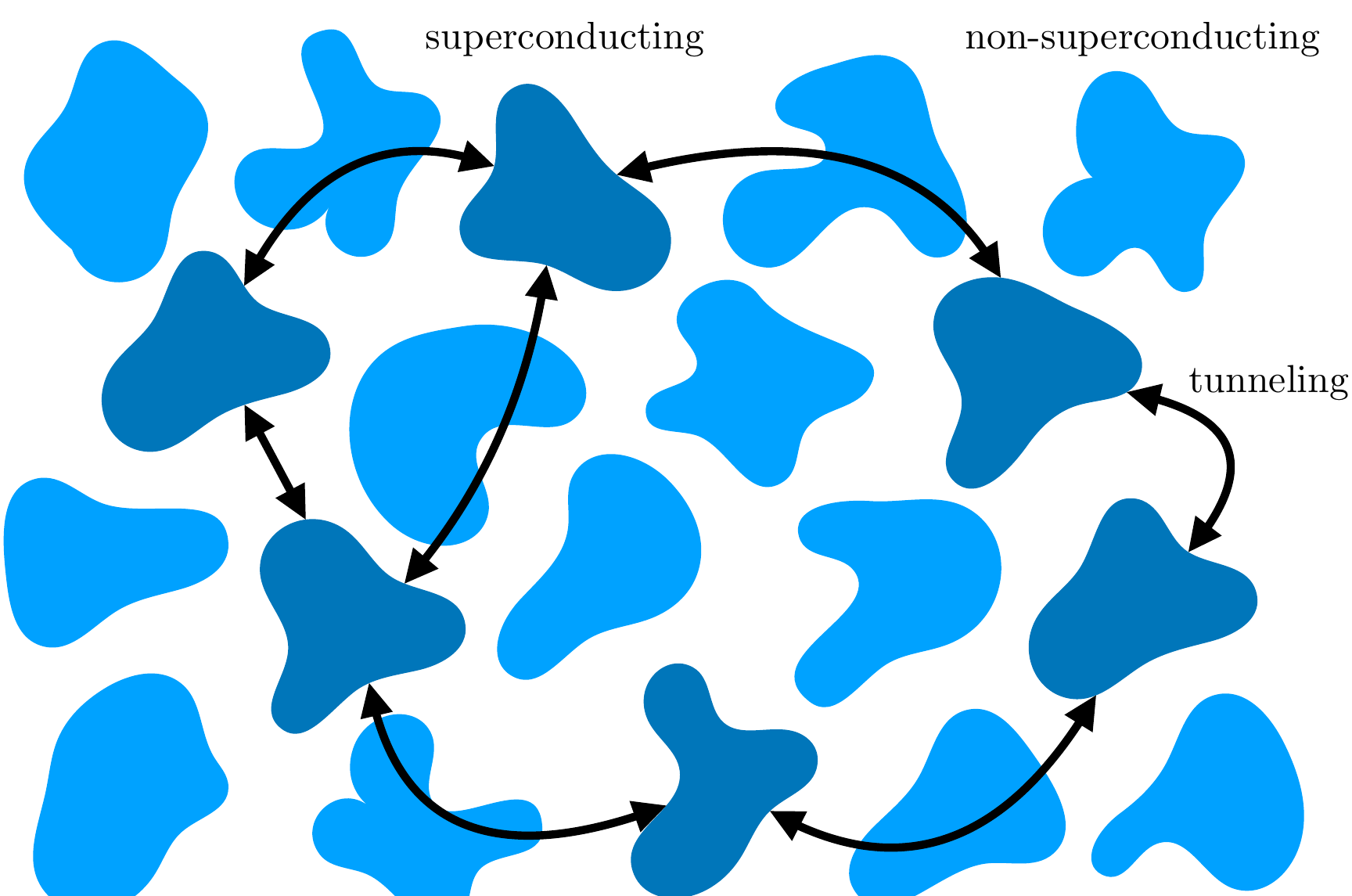}
\caption{A schematic illustration of a superconducting Josephson network of superconducting grains with random geometries. Dark and light blue regions represent superconducting and non-superconducting grains respectively. Arrows represent Josephson tunneling between the grains. }\label{Fig3}
\end{figure}

\begin{acknowledgments}
The authors thank Andy Millis, Amit Vikram, and Masoud Mohammadi for useful discussions. This work was supported by U.S. Department of Energy, Office of Science, Basic Energy Sciences under Award No. DE-SC0001911 (analytical random matrix theory by V.G. and A.G.) and DARPA HR00112490310 (numerical simulations). 
\end{acknowledgments}

\bibliographystyle{apsrev4-1}
\bibliography{bibl}

%merlin.mbs apsrev4-1.bst 2010-07-25 4.21a (PWD, AO, DPC) hacked
%Control: key (0)
%Control: author (72) initials jnrlst
%Control: editor formatted (1) identically to author
%Control: production of article title (-1) disabled
%Control: page (0) single
%Control: year (1) truncated
%Control: production of eprint (0) enabled
\begin{thebibliography}{30}%
\makeatletter
\providecommand \@ifxundefined [1]{%
 \@ifx{#1\undefined}
}%
\providecommand \@ifnum [1]{%
 \ifnum #1\expandafter \@firstoftwo
 \else \expandafter \@secondoftwo
 \fi
}%
\providecommand \@ifx [1]{%
 \ifx #1\expandafter \@firstoftwo
 \else \expandafter \@secondoftwo
 \fi
}%
\providecommand \natexlab [1]{#1}%
\providecommand \enquote  [1]{``#1''}%
\providecommand \bibnamefont  [1]{#1}%
\providecommand \bibfnamefont [1]{#1}%
\providecommand \citenamefont [1]{#1}%
\providecommand \href@noop [0]{\@secondoftwo}%
\providecommand \href [0]{\begingroup \@sanitize@url \@href}%
\providecommand \@href[1]{\@@startlink{#1}\@@href}%
\providecommand \@@href[1]{\endgroup#1\@@endlink}%
\providecommand \@sanitize@url [0]{\catcode `\\12\catcode `\$12\catcode
  `\&12\catcode `\#12\catcode `\^12\catcode `\_12\catcode `\%12\relax}%
\providecommand \@@startlink[1]{}%
\providecommand \@@endlink[0]{}%
\providecommand \url  [0]{\begingroup\@sanitize@url \@url }%
\providecommand \@url [1]{\endgroup\@href {#1}{\urlprefix }}%
\providecommand \urlprefix  [0]{URL }%
\providecommand \Eprint [0]{\href }%
\providecommand \doibase [0]{http://dx.doi.org/}%
\providecommand \selectlanguage [0]{\@gobble}%
\providecommand \bibinfo  [0]{\@secondoftwo}%
\providecommand \bibfield  [0]{\@secondoftwo}%
\providecommand \translation [1]{[#1]}%
\providecommand \BibitemOpen [0]{}%
\providecommand \bibitemStop [0]{}%
\providecommand \bibitemNoStop [0]{.\EOS\space}%
\providecommand \EOS [0]{\spacefactor3000\relax}%
\providecommand \BibitemShut  [1]{\csname bibitem#1\endcsname}%
\let\auto@bib@innerbib\@empty
%</preamble>
\bibitem [{\citenamefont {Abeles}\ \emph {et~al.}(1966)\citenamefont {Abeles},
  \citenamefont {Cohen},\ and\ \citenamefont {Cullen}}]{abeles1966enhancement}%
  \BibitemOpen
  \bibfield  {author} {\bibinfo {author} {\bibfnamefont {B.}~\bibnamefont
  {Abeles}}, \bibinfo {author} {\bibfnamefont {R.~W.}\ \bibnamefont {Cohen}}, \
  and\ \bibinfo {author} {\bibfnamefont {G.}~\bibnamefont {Cullen}},\
  }\href@noop {} {\bibfield  {journal} {\bibinfo  {journal} {Physical Review
  Letters}\ }\textbf {\bibinfo {volume} {17}},\ \bibinfo {pages} {632}
  (\bibinfo {year} {1966})}\BibitemShut {NoStop}%
\bibitem [{\citenamefont {Thomas}\ \emph {et~al.}(2019)\citenamefont {Thomas},
  \citenamefont {Devaux}, \citenamefont {Nagarajan}, \citenamefont {Chervy},
  \citenamefont {Seidel}, \citenamefont {Hagenm{\"u}ller}, \citenamefont
  {Sch{\"u}tz}, \citenamefont {Schachenmayer}, \citenamefont {Genet},
  \citenamefont {Pupillo} \emph {et~al.}}]{thomas2019exploring}%
  \BibitemOpen
  \bibfield  {author} {\bibinfo {author} {\bibfnamefont {A.}~\bibnamefont
  {Thomas}}, \bibinfo {author} {\bibfnamefont {E.}~\bibnamefont {Devaux}},
  \bibinfo {author} {\bibfnamefont {K.}~\bibnamefont {Nagarajan}}, \bibinfo
  {author} {\bibfnamefont {T.}~\bibnamefont {Chervy}}, \bibinfo {author}
  {\bibfnamefont {M.}~\bibnamefont {Seidel}}, \bibinfo {author} {\bibfnamefont
  {D.}~\bibnamefont {Hagenm{\"u}ller}}, \bibinfo {author} {\bibfnamefont
  {S.}~\bibnamefont {Sch{\"u}tz}}, \bibinfo {author} {\bibfnamefont
  {J.}~\bibnamefont {Schachenmayer}}, \bibinfo {author} {\bibfnamefont
  {C.}~\bibnamefont {Genet}}, \bibinfo {author} {\bibfnamefont
  {G.}~\bibnamefont {Pupillo}},  \emph {et~al.},\ }\href@noop {} {\bibfield
  {journal} {\bibinfo  {journal} {arXiv preprint arXiv:1911.01459}\ } (\bibinfo
  {year} {2019})}\BibitemShut {NoStop}%
\bibitem [{\citenamefont {Smolyaninova}\ \emph {et~al.}(2016)\citenamefont
  {Smolyaninova}, \citenamefont {Jensen}, \citenamefont {Zimmerman},
  \citenamefont {Prestigiacomo}, \citenamefont {Osofsky}, \citenamefont {Kim},
  \citenamefont {Bassim}, \citenamefont {Xing}, \citenamefont {Qazilbash},\
  and\ \citenamefont {Smolyaninov}}]{smolyaninova2016enhanced}%
  \BibitemOpen
  \bibfield  {author} {\bibinfo {author} {\bibfnamefont {V.~N.}\ \bibnamefont
  {Smolyaninova}}, \bibinfo {author} {\bibfnamefont {C.}~\bibnamefont
  {Jensen}}, \bibinfo {author} {\bibfnamefont {W.}~\bibnamefont {Zimmerman}},
  \bibinfo {author} {\bibfnamefont {J.~C.}\ \bibnamefont {Prestigiacomo}},
  \bibinfo {author} {\bibfnamefont {M.~S.}\ \bibnamefont {Osofsky}}, \bibinfo
  {author} {\bibfnamefont {H.}~\bibnamefont {Kim}}, \bibinfo {author}
  {\bibfnamefont {N.}~\bibnamefont {Bassim}}, \bibinfo {author} {\bibfnamefont
  {Z.}~\bibnamefont {Xing}}, \bibinfo {author} {\bibfnamefont {M.~M.}\
  \bibnamefont {Qazilbash}}, \ and\ \bibinfo {author} {\bibfnamefont {I.~I.}\
  \bibnamefont {Smolyaninov}},\ }\href@noop {} {\bibfield  {journal} {\bibinfo
  {journal} {Scientific reports}\ }\textbf {\bibinfo {volume} {6}},\ \bibinfo
  {pages} {34140} (\bibinfo {year} {2016})}\BibitemShut {NoStop}%
\bibitem [{\citenamefont {Cohen}\ and\ \citenamefont
  {Abeles}(1968)}]{cohen1968superconductivity}%
  \BibitemOpen
  \bibfield  {author} {\bibinfo {author} {\bibfnamefont {R.~W.}\ \bibnamefont
  {Cohen}}\ and\ \bibinfo {author} {\bibfnamefont {B.}~\bibnamefont {Abeles}},\
  }\href@noop {} {\bibfield  {journal} {\bibinfo  {journal} {Physical Review}\
  }\textbf {\bibinfo {volume} {168}},\ \bibinfo {pages} {444} (\bibinfo {year}
  {1968})}\BibitemShut {NoStop}%
\bibitem [{\citenamefont {Strongin}\ \emph {et~al.}(1968)\citenamefont
  {Strongin}, \citenamefont {Kammerer}, \citenamefont {Crow}, \citenamefont
  {Parks}, \citenamefont {Douglass~Jr},\ and\ \citenamefont
  {Jensen}}]{strongin1968enhanced}%
  \BibitemOpen
  \bibfield  {author} {\bibinfo {author} {\bibfnamefont {M.}~\bibnamefont
  {Strongin}}, \bibinfo {author} {\bibfnamefont {O.}~\bibnamefont {Kammerer}},
  \bibinfo {author} {\bibfnamefont {J.}~\bibnamefont {Crow}}, \bibinfo {author}
  {\bibfnamefont {R.}~\bibnamefont {Parks}}, \bibinfo {author} {\bibfnamefont
  {D.}~\bibnamefont {Douglass~Jr}}, \ and\ \bibinfo {author} {\bibfnamefont
  {M.}~\bibnamefont {Jensen}},\ }\href@noop {} {\bibfield  {journal} {\bibinfo
  {journal} {Physical Review Letters}\ }\textbf {\bibinfo {volume} {21}},\
  \bibinfo {pages} {1320} (\bibinfo {year} {1968})}\BibitemShut {NoStop}%
\bibitem [{\citenamefont {Prischepa}\ and\ \citenamefont
  {Kushnir}(2023)}]{prischepa2023phonon}%
  \BibitemOpen
  \bibfield  {author} {\bibinfo {author} {\bibfnamefont {S.}~\bibnamefont
  {Prischepa}}\ and\ \bibinfo {author} {\bibfnamefont {V.}~\bibnamefont
  {Kushnir}},\ }\href@noop {} {\bibfield  {journal} {\bibinfo  {journal}
  {Journal of Physics: Condensed Matter}\ }\textbf {\bibinfo {volume} {35}},\
  \bibinfo {pages} {313003} (\bibinfo {year} {2023})}\BibitemShut {NoStop}%
\bibitem [{\citenamefont {Garc{\'\i}a-Garc{\'\i}a}\ \emph
  {et~al.}(2011)\citenamefont {Garc{\'\i}a-Garc{\'\i}a}, \citenamefont
  {Urbina}, \citenamefont {Yuzbashyan}, \citenamefont {Richter},\ and\
  \citenamefont {Altshuler}}]{garcia2011bcs}%
  \BibitemOpen
  \bibfield  {author} {\bibinfo {author} {\bibfnamefont {A.~M.}\ \bibnamefont
  {Garc{\'\i}a-Garc{\'\i}a}}, \bibinfo {author} {\bibfnamefont {J.~D.}\
  \bibnamefont {Urbina}}, \bibinfo {author} {\bibfnamefont {E.~A.}\
  \bibnamefont {Yuzbashyan}}, \bibinfo {author} {\bibfnamefont
  {K.}~\bibnamefont {Richter}}, \ and\ \bibinfo {author} {\bibfnamefont
  {B.~L.}\ \bibnamefont {Altshuler}},\ }\href@noop {} {\bibfield  {journal}
  {\bibinfo  {journal} {Physical Review B---Condensed Matter and Materials
  Physics}\ }\textbf {\bibinfo {volume} {83}},\ \bibinfo {pages} {014510}
  (\bibinfo {year} {2011})}\BibitemShut {NoStop}%
\bibitem [{\citenamefont {Garc{\'\i}a-Garc{\'\i}a}\ \emph
  {et~al.}(2008)\citenamefont {Garc{\'\i}a-Garc{\'\i}a}, \citenamefont
  {Urbina}, \citenamefont {Yuzbashyan}, \citenamefont {Richter},\ and\
  \citenamefont {Altshuler}}]{garcia2008bardeen}%
  \BibitemOpen
  \bibfield  {author} {\bibinfo {author} {\bibfnamefont {A.~M.}\ \bibnamefont
  {Garc{\'\i}a-Garc{\'\i}a}}, \bibinfo {author} {\bibfnamefont {J.~D.}\
  \bibnamefont {Urbina}}, \bibinfo {author} {\bibfnamefont {E.~A.}\
  \bibnamefont {Yuzbashyan}}, \bibinfo {author} {\bibfnamefont
  {K.}~\bibnamefont {Richter}}, \ and\ \bibinfo {author} {\bibfnamefont
  {B.~L.}\ \bibnamefont {Altshuler}},\ }\href@noop {} {\bibfield  {journal}
  {\bibinfo  {journal} {Physical review letters}\ }\textbf {\bibinfo {volume}
  {100}},\ \bibinfo {pages} {187001} (\bibinfo {year} {2008})}\BibitemShut
  {NoStop}%
\bibitem [{\citenamefont {Tanner}\ and\ \citenamefont
  {S{\o}ndergaard}(2007)}]{tanner2007wave}%
  \BibitemOpen
  \bibfield  {author} {\bibinfo {author} {\bibfnamefont {G.}~\bibnamefont
  {Tanner}}\ and\ \bibinfo {author} {\bibfnamefont {N.}~\bibnamefont
  {S{\o}ndergaard}},\ }\href@noop {} {\bibfield  {journal} {\bibinfo  {journal}
  {Journal of Physics A: Mathematical and Theoretical}\ }\textbf {\bibinfo
  {volume} {40}},\ \bibinfo {pages} {R443} (\bibinfo {year}
  {2007})}\BibitemShut {NoStop}%
\bibitem [{\citenamefont {Animalu}\ \emph {et~al.}(1966)\citenamefont
  {Animalu}, \citenamefont {Bonsignori},\ and\ \citenamefont
  {Bortolani}}]{animalu1966phonon}%
  \BibitemOpen
  \bibfield  {author} {\bibinfo {author} {\bibfnamefont {A.~O.~E.}\
  \bibnamefont {Animalu}}, \bibinfo {author} {\bibfnamefont {F.}~\bibnamefont
  {Bonsignori}}, \ and\ \bibinfo {author} {\bibfnamefont {V.}~\bibnamefont
  {Bortolani}},\ }\href {\doibase 10.1007/BF02710433} {\bibfield  {journal}
  {\bibinfo  {journal} {Il Nuovo Cimento B (1965-1970)}\ }\textbf {\bibinfo
  {volume} {44}},\ \bibinfo {pages} {159} (\bibinfo {year} {1966})}\BibitemShut
  {NoStop}%
\bibitem [{\citenamefont {Achenbach}\ \emph {et~al.}(1982)\citenamefont
  {Achenbach}, \citenamefont {Gautesen},\ and\ \citenamefont
  {McMaken}}]{achenbach1982ray}%
  \BibitemOpen
  \bibfield  {author} {\bibinfo {author} {\bibfnamefont {J.}~\bibnamefont
  {Achenbach}}, \bibinfo {author} {\bibfnamefont {A.}~\bibnamefont {Gautesen}},
  \ and\ \bibinfo {author} {\bibfnamefont {H.}~\bibnamefont {McMaken}},\
  }\href@noop {} {{\selectlanguage {English}\emph {\bibinfo {title} {Ray
  Methods for Waves in Elastic Solids: with Applications to Scattering by
  Cracks}}}}\ (\bibinfo  {publisher} {Pitman Advanced Publishing Program},\
  \bibinfo {year} {1982})\BibitemShut {NoStop}%
\bibitem [{\citenamefont {Landau}\ \emph {et~al.}(2012)\citenamefont {Landau},
  \citenamefont {Pitaevskii}, \citenamefont {Kosevich},\ and\ \citenamefont
  {Lifshitz}}]{landau2012theory}%
  \BibitemOpen
  \bibfield  {author} {\bibinfo {author} {\bibfnamefont {L.~D.}\ \bibnamefont
  {Landau}}, \bibinfo {author} {\bibfnamefont {L.}~\bibnamefont {Pitaevskii}},
  \bibinfo {author} {\bibfnamefont {A.~M.}\ \bibnamefont {Kosevich}}, \ and\
  \bibinfo {author} {\bibfnamefont {E.~M.}\ \bibnamefont {Lifshitz}},\
  }\href@noop {} {\emph {\bibinfo {title} {Theory of elasticity: volume 7}}},\
  Vol.~\bibinfo {volume} {7}\ (\bibinfo  {publisher} {Elsevier},\ \bibinfo
  {year} {2012})\BibitemShut {NoStop}%
\bibitem [{\citenamefont {Weyl}(1911)}]{weyl1911asymptotische}%
  \BibitemOpen
  \bibfield  {author} {\bibinfo {author} {\bibfnamefont {H.}~\bibnamefont
  {Weyl}},\ }\href@noop {} {\bibfield  {journal} {\bibinfo  {journal}
  {Nachrichten von der Gesellschaft der Wissenschaften zu G{\"o}ttingen,
  Mathematisch-Physikalische Klasse}\ }\textbf {\bibinfo {volume} {1911}},\
  \bibinfo {pages} {110} (\bibinfo {year} {1911})}\BibitemShut {NoStop}%
\bibitem [{\citenamefont {Bertelsen}\ \emph {et~al.}(2000)\citenamefont
  {Bertelsen}, \citenamefont {Ellegaard},\ and\ \citenamefont
  {Hugues}}]{bertelsen2000distribution}%
  \BibitemOpen
  \bibfield  {author} {\bibinfo {author} {\bibfnamefont {P.}~\bibnamefont
  {Bertelsen}}, \bibinfo {author} {\bibfnamefont {C.}~\bibnamefont
  {Ellegaard}}, \ and\ \bibinfo {author} {\bibfnamefont {E.}~\bibnamefont
  {Hugues}},\ }\href@noop {} {\bibfield  {journal} {\bibinfo  {journal} {The
  European Physical Journal B-Condensed Matter and Complex Systems}\ }\textbf
  {\bibinfo {volume} {15}},\ \bibinfo {pages} {87} (\bibinfo {year}
  {2000})}\BibitemShut {NoStop}%
\bibitem [{\citenamefont {Fassbender}\ \emph {et~al.}(1989)\citenamefont
  {Fassbender}, \citenamefont {Hoffmann},\ and\ \citenamefont
  {Arnold}}]{fassbender1989efficient}%
  \BibitemOpen
  \bibfield  {author} {\bibinfo {author} {\bibfnamefont {S.}~\bibnamefont
  {Fassbender}}, \bibinfo {author} {\bibfnamefont {B.}~\bibnamefont
  {Hoffmann}}, \ and\ \bibinfo {author} {\bibfnamefont {W.}~\bibnamefont
  {Arnold}},\ }\href@noop {} {\bibfield  {journal} {\bibinfo  {journal}
  {Materials Science and Engineering: A}\ }\textbf {\bibinfo {volume} {122}},\
  \bibinfo {pages} {37} (\bibinfo {year} {1989})}\BibitemShut {NoStop}%
\bibitem [{\citenamefont {David}\ \emph {et~al.}(1963)\citenamefont {David},
  \citenamefont {Van~der Laan},\ and\ \citenamefont
  {Poulis}}]{david1963attenuation}%
  \BibitemOpen
  \bibfield  {author} {\bibinfo {author} {\bibfnamefont {R.}~\bibnamefont
  {David}}, \bibinfo {author} {\bibfnamefont {H.}~\bibnamefont {Van~der Laan}},
  \ and\ \bibinfo {author} {\bibfnamefont {N.}~\bibnamefont {Poulis}},\
  }\href@noop {} {\bibfield  {journal} {\bibinfo  {journal} {Physica}\ }\textbf
  {\bibinfo {volume} {29}},\ \bibinfo {pages} {357} (\bibinfo {year}
  {1963})}\BibitemShut {NoStop}%
\bibitem [{\citenamefont {Lide}(2004)}]{lide2004crc}%
  \BibitemOpen
  \bibfield  {author} {\bibinfo {author} {\bibfnamefont {D.~R.}\ \bibnamefont
  {Lide}},\ }\href@noop {} {\emph {\bibinfo {title} {CRC handbook of chemistry
  and physics}}},\ Vol.~\bibinfo {volume} {85}\ (\bibinfo  {publisher} {CRC
  press},\ \bibinfo {year} {2004})\BibitemShut {NoStop}%
\bibitem [{\citenamefont {Baratta}\ \emph {et~al.}(2023)\citenamefont
  {Baratta}, \citenamefont {Dean}, \citenamefont {Dokken}, \citenamefont
  {Habera}, \citenamefont {Hale}, \citenamefont {Richardson}, \citenamefont
  {Rognes}, \citenamefont {Scroggs}, \citenamefont {Sime},\ and\ \citenamefont
  {Wells}}]{barrata2023dolfinx}%
  \BibitemOpen
  \bibfield  {author} {\bibinfo {author} {\bibfnamefont {I.~A.}\ \bibnamefont
  {Baratta}}, \bibinfo {author} {\bibfnamefont {J.~P.}\ \bibnamefont {Dean}},
  \bibinfo {author} {\bibfnamefont {J.~S.}\ \bibnamefont {Dokken}}, \bibinfo
  {author} {\bibfnamefont {M.}~\bibnamefont {Habera}}, \bibinfo {author}
  {\bibfnamefont {J.~S.}\ \bibnamefont {Hale}}, \bibinfo {author}
  {\bibfnamefont {C.~N.}\ \bibnamefont {Richardson}}, \bibinfo {author}
  {\bibfnamefont {M.~E.}\ \bibnamefont {Rognes}}, \bibinfo {author}
  {\bibfnamefont {M.~W.}\ \bibnamefont {Scroggs}}, \bibinfo {author}
  {\bibfnamefont {N.}~\bibnamefont {Sime}}, \ and\ \bibinfo {author}
  {\bibfnamefont {G.~N.}\ \bibnamefont {Wells}},\ }\href {\doibase
  10.5281/zenodo.10447666} {\enquote {\bibinfo {title} {{DOLFINx: The next
  generation FEniCS problem solving environment}},}\ } (\bibinfo {year}
  {2023})\BibitemShut {NoStop}%
\bibitem [{\citenamefont {Marsiglio}(2020)}]{marsiglio2020eliashberg}%
  \BibitemOpen
  \bibfield  {author} {\bibinfo {author} {\bibfnamefont {F.}~\bibnamefont
  {Marsiglio}},\ }\href@noop {} {\bibfield  {journal} {\bibinfo  {journal}
  {Annals of Physics}\ }\textbf {\bibinfo {volume} {417}},\ \bibinfo {pages}
  {168102} (\bibinfo {year} {2020})}\BibitemShut {NoStop}%
\bibitem [{\citenamefont {McMillan}(1968)}]{mcmillan1968transition}%
  \BibitemOpen
  \bibfield  {author} {\bibinfo {author} {\bibfnamefont {W.}~\bibnamefont
  {McMillan}},\ }\href@noop {} {\bibfield  {journal} {\bibinfo  {journal}
  {Physical Review}\ }\textbf {\bibinfo {volume} {167}},\ \bibinfo {pages}
  {331} (\bibinfo {year} {1968})}\BibitemShut {NoStop}%
\bibitem [{\citenamefont {Allen}\ and\ \citenamefont
  {Dynes}(1975)}]{allen1975transition}%
  \BibitemOpen
  \bibfield  {author} {\bibinfo {author} {\bibfnamefont {P.~B.}\ \bibnamefont
  {Allen}}\ and\ \bibinfo {author} {\bibfnamefont {R.}~\bibnamefont {Dynes}},\
  }\href@noop {} {\bibfield  {journal} {\bibinfo  {journal} {Physical Review
  B}\ }\textbf {\bibinfo {volume} {12}},\ \bibinfo {pages} {905} (\bibinfo
  {year} {1975})}\BibitemShut {NoStop}%
\bibitem [{\citenamefont {Ummarino}\ \emph {et~al.}(2013)\citenamefont
  {Ummarino} \emph {et~al.}}]{ummarino2013eliashberg}%
  \BibitemOpen
  \bibfield  {author} {\bibinfo {author} {\bibfnamefont {G.}~\bibnamefont
  {Ummarino}} \emph {et~al.},\ }\href@noop {} {\bibfield  {journal} {\bibinfo
  {journal} {Emergent Phenomena in Correlated Matter}\ }\textbf {\bibinfo
  {volume} {3}},\ \bibinfo {pages} {13} (\bibinfo {year} {2013})}\BibitemShut
  {NoStop}%
\bibitem [{\citenamefont {Berry}(1988)}]{berry1988semiclassical}%
  \BibitemOpen
  \bibfield  {author} {\bibinfo {author} {\bibfnamefont {M.~V.}\ \bibnamefont
  {Berry}},\ }\href {\doibase 10.1088/0951-7715/1/3/001} {\bibfield  {journal}
  {\bibinfo  {journal} {Nonlinearity}\ }\textbf {\bibinfo {volume} {1}},\
  \bibinfo {pages} {399} (\bibinfo {year} {1988})}\BibitemShut {NoStop}%
\bibitem [{\citenamefont {Akolzin}\ and\ \citenamefont
  {Weaver}(2004)}]{akolzin2004generalized}%
  \BibitemOpen
  \bibfield  {author} {\bibinfo {author} {\bibfnamefont {A.}~\bibnamefont
  {Akolzin}}\ and\ \bibinfo {author} {\bibfnamefont {R.~L.}\ \bibnamefont
  {Weaver}},\ }\href@noop {} {\bibfield  {journal} {\bibinfo  {journal}
  {Physical Review E---Statistical, Nonlinear, and Soft Matter Physics}\
  }\textbf {\bibinfo {volume} {70}},\ \bibinfo {pages} {046212} (\bibinfo
  {year} {2004})}\BibitemShut {NoStop}%
\bibitem [{\citenamefont {Galitski}\ and\ \citenamefont
  {Larkin}(2001)}]{galitski2001disorder}%
  \BibitemOpen
  \bibfield  {author} {\bibinfo {author} {\bibfnamefont {V.}~\bibnamefont
  {Galitski}}\ and\ \bibinfo {author} {\bibfnamefont {A.}~\bibnamefont
  {Larkin}},\ }\href@noop {} {\bibfield  {journal} {\bibinfo  {journal}
  {Physical Review Letters}\ }\textbf {\bibinfo {volume} {87}},\ \bibinfo
  {pages} {087001} (\bibinfo {year} {2001})}\BibitemShut {NoStop}%
\bibitem [{\citenamefont {Galitski}(2008)}]{galitski2008mesoscopic}%
  \BibitemOpen
  \bibfield  {author} {\bibinfo {author} {\bibfnamefont {V.}~\bibnamefont
  {Galitski}},\ }\href {\doibase 10.1103/PhysRevB.77.100502} {\bibfield
  {journal} {\bibinfo  {journal} {Phys. Rev. B}\ }\textbf {\bibinfo {volume}
  {77}},\ \bibinfo {pages} {100502} (\bibinfo {year} {2008})}\BibitemShut
  {NoStop}%
\bibitem [{\citenamefont {Kuhl}\ \emph {et~al.}(2005)\citenamefont {Kuhl},
  \citenamefont {St{\"o}ckmann},\ and\ \citenamefont
  {Weaver}}]{kuhl2005classical}%
  \BibitemOpen
  \bibfield  {author} {\bibinfo {author} {\bibfnamefont {U.}~\bibnamefont
  {Kuhl}}, \bibinfo {author} {\bibfnamefont {H.}~\bibnamefont {St{\"o}ckmann}},
  \ and\ \bibinfo {author} {\bibfnamefont {R.}~\bibnamefont {Weaver}},\
  }\href@noop {} {\bibfield  {journal} {\bibinfo  {journal} {Journal of Physics
  A: Mathematical and General}\ }\textbf {\bibinfo {volume} {38}},\ \bibinfo
  {pages} {10433} (\bibinfo {year} {2005})}\BibitemShut {NoStop}%
\bibitem [{\citenamefont {Belzig}\ \emph {et~al.}(1996)\citenamefont {Belzig},
  \citenamefont {Bruder},\ and\ \citenamefont {Sch{\"o}n}}]{belzig1996local}%
  \BibitemOpen
  \bibfield  {author} {\bibinfo {author} {\bibfnamefont {W.}~\bibnamefont
  {Belzig}}, \bibinfo {author} {\bibfnamefont {C.}~\bibnamefont {Bruder}}, \
  and\ \bibinfo {author} {\bibfnamefont {G.}~\bibnamefont {Sch{\"o}n}},\
  }\href@noop {} {\bibfield  {journal} {\bibinfo  {journal} {Physical Review
  B}\ }\textbf {\bibinfo {volume} {54}},\ \bibinfo {pages} {9443} (\bibinfo
  {year} {1996})}\BibitemShut {NoStop}%
\bibitem [{\citenamefont {Belzig}\ \emph {et~al.}(1999)\citenamefont {Belzig},
  \citenamefont {Wilhelm}, \citenamefont {Bruder}, \citenamefont {Sch{\"o}n},\
  and\ \citenamefont {Zaikin}}]{belzig1999quasiclassical}%
  \BibitemOpen
  \bibfield  {author} {\bibinfo {author} {\bibfnamefont {W.}~\bibnamefont
  {Belzig}}, \bibinfo {author} {\bibfnamefont {F.~K.}\ \bibnamefont {Wilhelm}},
  \bibinfo {author} {\bibfnamefont {C.}~\bibnamefont {Bruder}}, \bibinfo
  {author} {\bibfnamefont {G.}~\bibnamefont {Sch{\"o}n}}, \ and\ \bibinfo
  {author} {\bibfnamefont {A.~D.}\ \bibnamefont {Zaikin}},\ }\href@noop {}
  {\bibfield  {journal} {\bibinfo  {journal} {Superlattices and
  microstructures}\ }\textbf {\bibinfo {volume} {25}},\ \bibinfo {pages} {1251}
  (\bibinfo {year} {1999})}\BibitemShut {NoStop}%
\bibitem [{\citenamefont {Bergeret}\ \emph {et~al.}(2004)\citenamefont
  {Bergeret}, \citenamefont {Volkov},\ and\ \citenamefont
  {Efetov}}]{bergeret2004spin}%
  \BibitemOpen
  \bibfield  {author} {\bibinfo {author} {\bibfnamefont {F.~S.}\ \bibnamefont
  {Bergeret}}, \bibinfo {author} {\bibfnamefont {A.~F.}\ \bibnamefont
  {Volkov}}, \ and\ \bibinfo {author} {\bibfnamefont {K.~B.}\ \bibnamefont
  {Efetov}},\ }\href {\doibase 10.1209/epl/i2004-10003-3} {\bibfield  {journal}
  {\bibinfo  {journal} {Europhysics Letters}\ }\textbf {\bibinfo {volume}
  {66}},\ \bibinfo {pages} {111} (\bibinfo {year} {2004})}\BibitemShut
  {NoStop}%
\end{thebibliography}%

\appendix
%dummy comment inserted by tex2lyx to ensure that this paragraph is not empty%dummy comment inserted by tex2lyx to ensure that this paragraph is not empty

\section{Navier-Cauchy equations}

Here we review the elastic equations inside a grain. We define
the local lattice displacement vector $\vec{\phi}$ obeying \cite{tanner2007wave}:
\[
\rho\frac{d^{2}}{dt^{2}}\vec{\phi}=\nabla\cdot\hat{\sigma}(\vec{\phi}),
\]
where $\rho$ is the material density and $\sigma$ is the stress
tensor,

\[
\hat{\sigma}(\vec{\phi})=\xi{\bf 1}\text{Tr}\hat{\epsilon}+2\nu\hat{{\bf {\bf \epsilon}}}
\]
where $\lambda$ and $\mu$ are Lam\'e constants and the strain tensor
is assumed to be
\[
\epsilon_{i,j}=\frac{1}{2}\left(\partial_{j}\phi_{i}+\partial_{i}\phi_{j}\right)
\]
With this, we get:

\[
\rho\frac{d^{2}}{dt^{2}}\phi_{i}=\sum_{j}\partial_{j}\sigma_{j,i}(\vec{\phi})=\left(\xi+\nu\right)\partial_{i}\sum_{j}\partial_{j}\phi_{j}+\nu\sum_{j}\partial_{j}^{2}\phi_{i},
\]
or in the vector form:

\[
\rho\frac{d^{2}}{dt^{2}}\vec{\phi}=\left(\xi+\nu\right)\nabla\left(\nabla\cdot\vec{\phi}\right)+\nu\Delta\vec{\phi},
\]
The coefficients $\lambda$ and $\mu$ can be straightforwardly related to bulk sound velocities. Indeed,
in Fourier space we get:

\[
\omega^{2}\vec{\phi}_{{\bf k}}=\frac{\xi+\nu}{\rho}{\bf k}\left({\bf k}\cdot\vec{\phi}_{{\bf k}}\right)+\frac{\nu}{\rho}k^{2}\vec{\phi}_{{\bf k}}
\]
Projecting onto the longitudinal and transverse components we get:
$\omega=\sqrt{\frac{\xi+2\nu}{\rho}}k$ and $\omega=\sqrt{\nu/\rho}k$. The longitudinal
speed of sound is thus $c_{\parallel}=\sqrt{\frac{\xi+2\nu}{\rho}}$.

In our calculations, we assume the free-surface boundary conditions, which are equivalent
to the absence of restoring force at the boundary ${\bf n}\cdot\hat{\sigma}(\vec{\phi})=0$,
where ${\bf n}$ is the normal vector to the boundary.

\section{Derivation of the eigenvector ansatz}

Here we provide a heuristic derivation of the eigenvector ansatz Eq.~\eqref{eq:phonon_correlation}
following conventional argument related to the short-distance correlations
in chaotic systems. More precisely, we define: $f_{A}\left(\omega,{\bf r},{\bf r}'\right)=\sum_{l}C_{l}\left({\bf r}\right)C_{l}\left({\bf r}'\right)\delta\left(\omega-\omega_{l}\right)$,
where $C_{l}\left({\bf r}\right)$ is the eigenvector divergence.
The correlation function of divergences can now be inferred from
$f$ assuming an unbounded system as follows \cite{kuhl2005classical,bertelsen2000distribution,berry1988semiclassical}:

\begin{align*}
\overline{C_{l}\left({\bf r}\right)C_{l}\left({\bf r}'\right)} & =A^{-1}\nu_{\text{tot}}^{-1}\left(\omega\right)f_{\infty}\left(\omega\left|{\bf r}-{\bf r}'\right|\right),
\end{align*}
where $f_{\infty}=\lim_{A\rightarrow\infty}f_{A}$. 
\begin{align*}
f_{\infty}\left(\omega,{\bf r},{\bf r}'\right) & =\int\frac{d^{2}{\bf k}}{\left(2\pi\right)^{2}}k^{2}e^{i{\bf k}{\bf r}}\delta\left(\omega-c_{\parallel}k\right)\\
 & =\nu_{\parallel}\left(\omega\right)\frac{\omega^{2}}{c_{\parallel}^{2}}J_{0}\left(\frac{\omega}{c}\left|{\bf r}-{\bf r}'\right|\right)
\end{align*}
where $J_{0}$ is Bessel's function of the first kind and $\nu_{\parallel}\left(\omega\right)$
is the longitudinal density of states. %We note that $\tilde{\nu}_{\parallel}$
% is not necessary equivalent to the longitudinal DOS which can be obtained
% in a naive way (which we also investigate in Fig.~1~(a)):

% \[
% \nu_{\parallel}\left(\omega\right)=\int\frac{d^{2}{\bf k}}{\left(2\pi\right)^{2}}\sum_{l}\left|\frac{{\bf k}}{k}\cdot\vec{\phi}_{l}\left({\bf k}\right)\right|^{2}\delta\left(\omega-\omega_{l}\right).
% \]
% where $\vec{\phi}\left({\bf k}\right)=\int_{A}e^{-i{\bf kr}}\vec{\phi}\left({\bf r}\right)$.
% The discrepancy comes from the formal discontinuity of divergencies
% if they are computed in momentum space as ${\bf k}\cdot\vec{\phi}_{l}\left({\bf k}\right)$.

\section{Weyl's law for phonon billiards with free-surface boundary conditions}

Here we provide an explicit expression for the Weyl parameter, $\eta$, used in
Eq.~\eqref{eq:Nw}. This parameter was derived in Ref.~\cite{bertelsen2000distribution} as follows
\begin{align}
 & \eta=\frac{4}{\sqrt{\gamma}}-3+\frac{1}{\kappa}\nonumber \\
 & +\frac{4}{\pi}\int_{1/\kappa}^{1}\arctan\left\{ \frac{\left(t^{2}-1\right)^{2}}{4t^{2}\sqrt{1-t^{2}}\sqrt{t^{2}-\frac{1}{\kappa^{2}}}}\right\} dt,
\end{align}
where $\kappa=c_{\parallel}/c_{\perp}$ and $\gamma$ is a solution
to the following equation belonging to the interval $]0,1[$: 
\begin{align}
\gamma^{3}-8\gamma^{2}+8\left(3-\frac{2}{\kappa^{2}}\right)\gamma-16\left(1-\frac{1}{\kappa^{2}}\right)=0
\end{align}
We note that the value of $\eta$ is agreement with our numerical
simulations. 

\section{Diffusive limit}

Here we derive the effective interaction within the quasi-classical
approximation \cite{belzig1996local,belzig1999quasiclassical,bergeret2004spin} assuming
the interaction is changing sufficiently slowly in space which is the case in our parameter regime $k_{D}<k_{F}$. We start by rewriting
the self-energy equation in momentum space for the relative coordinate:

\begin{align*}
 & \hat{\Sigma}_{{\bf R},{\bf k}}\left(i\epsilon_{n}\right)=\\
 & -\frac{1}{\beta A}\sum_{n'}\sum_{{\bf k}'}{\cal D}_{{\bf R},{\bf k-k'}}\left(i\epsilon_{n}-i\epsilon_{n'}\right)\hat{\tau}_{3}\hat{{\cal G}}_{{\bf R},{\bf k'}}\left(i\epsilon_{n'}\right)\hat{\tau}_{3},\\
\end{align*}
where we performed the Fourier transform with respect to the relative
coordinate. The center-of-mass (COM) coordinate is defined as ${\bf R}=({\bf r}+{\bf r}')/2$).
In the following, we  perform a quasi-local approximation for the
phonon propagator by restricting both momenta
to the Fermi surface \cite{mcmillan1968transition,allen1975transition}.
In this case, the self-energy depends only on the COM coordinate.
Disorder scattering can be added in a similar way as we discuss in
the main text. We now consider a quasiclassical approximation for
electrons by defining $\hat{g}_{{\bf R}}\equiv\frac{i}{\pi}\int d\xi_{k}\hat{\tau}_{3}\hat{G}_{{\bf R},{\bf k}}$, which obeys the Usadel equation in the diffusive limit:

\begin{equation}
D\nabla\left(\hat{g}_{{\bf R}}\nabla\hat{g}_{{\bf R}}\right)=-\left[\hat{g}_{{\bf R}},\epsilon_{n}\tau_{3}+i\hat{\Sigma}_{{\bf R}}\tau_{3}\right],\label{eq:gx}
\end{equation}
where $D=v_{F}^{2}\tau_{\rm{el}}/2$ is the diffusion coefficient, $\tau_{\rm{el}}$ denotes
the disorder scattering rate and we restricted the self-energy to
its value on the Fermi surface $\hat{\Sigma}_{{\bf R}}\left(i\epsilon_{n}\right)\equiv\hat{\Sigma}_{{\bf R},k=k_{F}}\left(i\epsilon_{n}\right)$
according to the local approximation. The vacuum boundary condition
for the quasiclassical Green's function is given by ${\bf {n}\cdot\nabla g_{{\bf R}}=0}$,
where ${\bf {n}}$ is the vector normal to the boundary. We also perform
the standard quasi-local approximation for the self-energy:

\[
\hat{\Sigma}_{{\bf R}}\left(i\epsilon_{n}\right)=\frac{1}{\beta}\nu_{0}^{F}i\pi\sum_{n'}\left\langle {\cal D}_{{\bf R},{\bf k_{F}-k'_{F}}}\left(i\epsilon_{n}-i\epsilon_{n'}\right)\right\rangle \hat{g}_{{\bf R}}\left(i\epsilon_{n'}\right)\hat{\tau}_{3},
\]
where the phonon propagator is averaged over direction of the difference
of two Fermi wavevectors ${\bf k_{F}-k'_{F}}$ and $\nu_{0}^{F}$
is the fermion density of states. We now look for a critical temperature
and linearize the Usadel's equation with respect to the anomalous
component. To this end we use the ansatz $g_{{\bf R}}=g^{(0)}+f_{{\bf R}}=\text{sign}\epsilon_{n}\tau_{3}+f_{{\bf R}}$, where $f_{{\bf R}}$ is purely off-diagonal. We write
the local self-energy in the conventional form $i\hat{\Sigma}_{{\bf R}}\tau_{3}\approx Z_{{\bf R}}\left(i\epsilon_{n}\right)\tau_{0}+\Delta_{{\bf R}}\left(i\epsilon_{n}\right)\tau_{1}$,
where $Z$ is the quasiparticle renormalization factor and $\Delta_{{\bf R}}$
is the gap. From Eq.~\eqref{eq:gx} we get:

\begin{equation}
D\nabla^{2}\hat{f}_{{\bf R}}=-2\tau_{1}\Delta_{{\bf R}}+2\left|\epsilon_{n}\right|Z_{{\bf R}}\left(i\epsilon_{n}\right)\hat{f}_{{\bf R}},\label{eq:gx-2-1-1}
\end{equation}
The characteristic length of this diffusion equation is thus given
by the coherence length of the superconductor $\xi\approx\sqrt{D/\left(2\pi T\overline{Z_{{\bf R}}}\right)}$
as expected. In the limit when the coherence length is large, the
$f_{{\bf R}}$ changes little between the boundaries and we can write

\[
\hat{f}_{{\bf R}}\approx\frac{\tau_{1}\overline{\Delta_{{\bf R}}}\left(i\epsilon_{n}\right)}{\left|\epsilon_{n}\right|\overline{Z_{{\bf R}}\left(i\epsilon_{n}\right)}}
\]
where the averaging is taken over the grain area. The self-consistency
equation becomes: 
\[
\Delta_{{\bf R}}\left(i\epsilon_{n}\right)=-\frac{1}{\beta}\nu_{0}\pi\sum_{n'}\left\langle {\cal D}_{{\bf R},{\bf k_{F}-k'_{F}}}\left(i\epsilon_{n}-i\epsilon_{n'}\right)\right\rangle \frac{\overline{\Delta_{{\bf R}}}\left(i\epsilon_{n'}\right)}{\left|\epsilon_{n}\right|\overline{Z_{{\bf R}}\left(i\epsilon_{n'}\right)}}.
\]
We note that $\Delta_{{\bf {R}}}$ is simply the self-energy and it
is not equivalent to the superconducting gap which is uniform. Moreover,
we are interested in averaging over the grain area (since it defines
$T_{c}$ and the actual gap):

\[
\overline{\Delta_{{\bf R}}}\left(i\epsilon_{n}\right)=-\frac{1}{\beta}\nu_{0}\pi\sum_{n'}\overline{\left\langle {\cal D}_{{\bf R},{\bf k_{F}-k'_{F}}}\left(i\epsilon_{n}-i\epsilon_{n'}\right)\right\rangle }\frac{\overline{\Delta_{{\bf R}}}\left(i\epsilon_{n'}\right)}{\left|\epsilon_{n}\right|\overline{Z_{{\bf R}}\left(i\epsilon_{n'}\right)}}.
\]
Let us now explicitly derive the interaction (note that integral can
be taken over the infinite space):

\begin{align*}
\overline{\left\langle {\cal D}_{{\bf R},{\bf k_{F}-k'_{F}}}\left(i\Omega_{n}\right)\right\rangle } & \equiv\int\frac{d^{2}{\bf R}}{A}{\cal D}_{{\bf R},{\bf k_{F}-k'_{F}}}\left(i\Omega_{n}\right)=\\
 & =\int\frac{d^{2}{\bf R}}{A}\int d^{2}{\bf r}\left\langle e^{-i\left({\bf k}_{\text{F}}-{\bf k}_{\text{F}}^{\prime}\right){\bf r}}\right\rangle {\cal D}_{{\bf R},{\bf r}}\left(i\Omega_{n}\right)\\
 & =A^{-1}\int d^{2}{\bf r}d^{2}{\bf r}'J_{0}^{2}\left(k_{F}|{\bf {r}-{\bf {r}'|}}\right){\cal D}_{{\bf r},{\bf r}'}\left(i\Omega_{n}\right)
\end{align*}
Which is the same interaction as in the main text. 

We now provide details of an analytical estimation of the transition
temperature in Eq.~\eqref{Tc}. We first compute the BCS pairing
strength and the logarithmic cut-off frequency exactly numerically
and with the approximate Eliashberg function Eq.~\eqref{eq:analytics}
as shown in Fig.~\eqref{Fig4}~(a). The non-universal low-energy
behavior is modeled as a sharp cut-off. We find a nearly perfect fits
with our analytical estimate of spectral density Eq.~\eqref{eq:analytics}.

\begin{figure}
\includegraphics[scale=0.6]{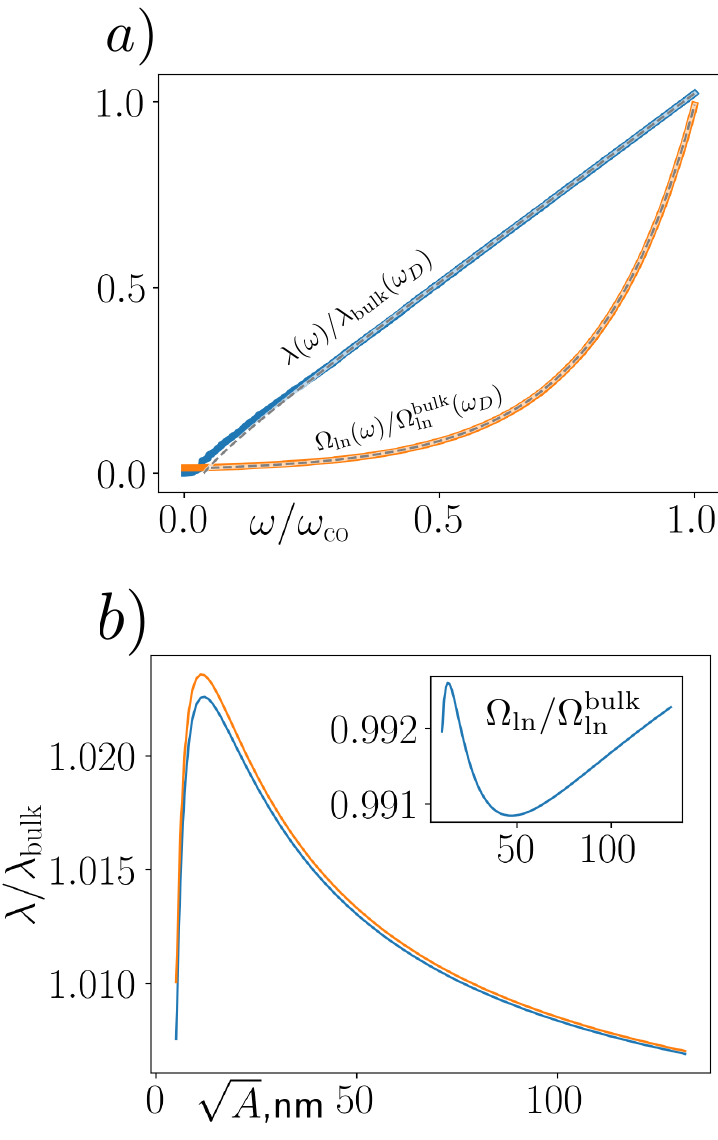}

\caption{Fit of the numerical data for $\lambda\left(\omega\right)$ and $\Omega_{ln}\left(\omega\right)$.
Grey dashed lines are fits using Eq.~\eqref{eq:analytics} with $\omega_{0}\approx1.1c_{\parallel}/\sqrt{A}$
for the BCS pairing strength and $\omega_{0}\approx1.5c_{\parallel}/\sqrt{A}$
for the logarithmic cut-off frequency.}

\label{Fig4}
\end{figure}
Let us now assume that we fix the grain shape but perform a scaling
transformation. We assume the low-energy cut-off frequencies are scaled
accordingly. At the same time, the high-energy behavior is correctly
captured by our analytical expression Eq.~\eqref{eq:analytics}.
Within these assumptions we find in the analytically tractable limit
$k_{F}\rightarrow\infty$ and expanding in the system size $P/A\sim L^{-1},$
$\omega_{0}\sim L^{-1}$:

\[
\frac{\lambda-\lambda_{\text{bulk}}}{\lambda_{\text{bulk}}}\approx\frac{\eta_{\parallel}P}{2Ak_{D}}\left\{ \log\left(\frac{\omega_{\text{D}}}{\omega_{0}e^{\chi}}\right)\right\} -\frac{\omega_{0}}{\omega_{\text{D}}}
\]
In Fig.~\eqref{Fig4}~(b) we compare this approximation with the
exact integral over Eliashberg function Eq.~\eqref{eq:analytics}.
We find a reasonably good agreement. We also numerically estimate
the change in $\Omega_{\text{ln}}$ which is found to be small and
we assume it can be absorbed into the frequency cut-off of the BCS
constant.

\subsection{Critical temperature}

We now discuss how the change in the BCS strength $\lambda$ and the
cut-off frequency $\Omega_{\text{ln}}$ affect the critical temperature.
From Eq.~\eqref{eq:MAD} we get

\[
\frac{\delta T_{c}}{T_{c}}\approx4.74\frac{\lambda-\lambda_{\text{bulk}}}{\lambda_{\text{bulk}}},
\]
where we used our estimation of the bulk pairing strength $\lambda_{\text{bulk}}\approx0.44$. 

\end{document}